\documentclass{UApreprint}
\usepackage{epsfig}
\usepackage{graphics}
\usepackage{fullpage}
\usepackage{verbatim}
\usepackage{amssymb}
\usepackage[tight]{subfigure}
\newcommand{\be}{\begin{equation}}
\newcommand{\ee}{\end{equation}}
\newcommand{\bea}{\begin{eqnarray}}
\newcommand{\eea}{\end{eqnarray}}
\newcommand{\bean}{\begin{eqnarray*}}
\newcommand{\eean}{\end{eqnarray*}}
\newcommand{\meg}{\mu \rightarrow e \gamma}
\newcommand{\half}{\frac{1}{2}}

\newcommand{\unit}{\mathbf{I}}
\newcommand{\tr}{\mathrm{Tr}}
\newcommand{\diag}{\mathrm{diag}}
\newcommand{\T}{^\mathrm{T}}
\newcommand{\Ye}{\mathbf{Y}_\mathrm{E}}
\newcommand{\Yn}{\mathbf{Y}_\mathrm{N}}
\newcommand{\Yd}{\mathbf{Y}_\mathrm{D}}
\newcommand{\Yu}{\mathbf{Y}_\mathrm{U}}
\newcommand{\Mnu}{\mathbf{M}_\mathrm{N}}
\newcommand{\Ae}{\mathbf{A}_\mathrm{E}}
\newcommand{\An}{\mathbf{A}_\mathrm{N}}
\newcommand{\Ad}{\mathbf{A}_\mathrm{D}}
\newcommand{\Au}{\mathbf{A}_\mathrm{U}}
\newcommand{\mls}{\mathbf{m}^2_{\tilde{\mathrm{L}}}}
\newcommand{\mes}{\mathbf{m}^2_{\tilde{\mathrm{E}}}}
\newcommand{\mns}{\mathbf{m}^2_{\tilde{\mathrm{N}}}}
\newcommand{\mqs}{\mathbf{m}^2_{\tilde{\mathrm{Q}}}}
\newcommand{\mds}{\mathbf{m}^2_{\tilde{\mathrm{D}}}}
\newcommand{\mus}{\mathbf{m}^2_{\tilde{\mathrm{U}}}}
\newcommand{\mhds}{m^2_\mathrm{H_d}}
\newcommand{\mhus}{m^2_\mathrm{H_u}}
\newcommand{\la}{\mathcal{L}}
\newcommand{\eb}{\mathbf{e}}
\newcommand{\nb}{\mathbf{\nu}}
\newcommand{\Lb}{\mathbf{L}}
\newcommand{\Eb}{\mathbf{E}}
\newcommand{\Nb}{\mathbf{N}}
\newcommand{\Ue}{\mathbf{U}_\mathrm{E}}
\newcommand{\Ve}{\mathbf{V}_\mathrm{E}}

\newcommand{\sw}{\sin \theta_\mathrm{W}}
\newcommand{\cw}{\cos \theta_\mathrm{W}}
\newcommand{\sB}{\sin \beta}
\newcommand{\cB}{\cos \beta}
\newcommand{\mz}{m_\mathrm{Z}}
\newcommand{\mw}{m_\mathrm{W}}
\newcommand{\MC}{\mathbf{M}_\mathrm{C}}
\newcommand{\OL}{\mathbf{O}_\mathrm{L}}
\newcommand{\OR}{\mathbf{O}_\mathrm{R}}
\newcommand{\Mne}{\mathbf{M}_\mathrm{ne}}
\newcommand{\ON}{\mathbf{O}_\mathrm{ne}}
\newcommand{\cc}{\;+\;\mathrm{c.\;c.}}
\newcommand{\Se}{\tilde{\mathbf{e}}}
\newcommand{\SE}{\tilde{\mathbf{E}}}

\newcommand{\mLL}{\mathbf{m}_\mathrm{LL}^2}
\newcommand{\mRL}{\mathbf{m}_\mathrm{RL}^2}
\newcommand{\mLR}{\mathbf{m}_\mathrm{RL}^{2\dagger}}
\newcommand{\mRR}{\mathbf{m}_\mathrm{RR}^2}
\newcommand{\ml}{\mathbf{m}_\mathrm{l}^2}
\newcommand{\mlfp}{\mathbf{m}_\mathrm{l}}
\newcommand{\Uf}{\mathbf{U}_\mathrm{\tilde{f}}}
\newcommand{\Un}{\mathbf{U}_\mathrm{\tilde{n}}}
\newcommand{\mnu}{\mathbf{m}_\nu}
\newcommand{\Hdz}{H_\mathrm{d}^0}
\newcommand{\Huz}{H_\mathrm{u}^0}
\newcommand{\Hdm}{H_\mathrm{d}^-}
\newcommand{\Hup}{H_\mathrm{u}^+}

\newcommand{\SB}{\tilde{B}}
\newcommand{\SWt}{\tilde{W}^3}
\newcommand{\SWp}{\tilde{W}^+}
\newcommand{\SWm}{\tilde{W}^-}
\newcommand{\SHdz}{\tilde{H}_\mathrm{d}^0}
\newcommand{\SHdm}{\tilde{H}_\mathrm{d}^-}
\newcommand{\SHup}{\tilde{H}_\mathrm{u}^+}
\newcommand{\SHuz}{\tilde{H}_\mathrm{u}^0}
\newcommand{\Msnu}{\mathbf{M}_{\tilde{\nu}}^2}
\newcommand{\Qb}{\mathbf{Q}}
\newcommand{\ub}{\mathbf{u}}
\newcommand{\db}{\mathbf{d}}
\newcommand{\Ub}{\mathbf{U}}
\newcommand{\Db}{\mathbf{D}}
\begin{document}
\title{Lepton Flavour Violation in a Class of Lopsided SO(10) Models}
\author{Ernest Jankowski\footnote{email: \texttt{ejankows@phys.ualberta.ca}}
and David W.\ Maybury\footnote{email: \texttt{dmaybury@phys.ualberta.ca}}}
\address{Department of Physics, University of Alberta, Edmonton  AB  T6G 2J1,
CANADA}
\date{}
\abstract{A class of predictive SO(10) grand unified
theories with highly asymmetric mass matrices, known as lopsided
textures, has been developed to accommodate the observed mixing in the
neutrino sector.
The model class effectively determines the rate for charged lepton flavour
violation, and in particular the branching ratio for $\meg$,
assuming that the supersymmetric GUT breaks directly to the constrained minimal
supersymmetric standard model (CMSSM). We
find that in light of the combined constraints on the CMSSM parameters from direct searches
and from the WMAP satellite observations, the resulting predicted
rate for $\meg$ in this model class can be within the current
experimental bounds for low $\tan \beta$, but that the next
generation of $\meg$ experiments would effectively rule out
this model class if LFV is not detected.}
\archive{hep-ph/0401132}
\preprintone{ALTA-TH-01-04}
\preprinttwo{}
\submit{}
\maketitle
\section{Introduction}
Neutrinos have been observed to oscillate between flavour states
\cite{Davis}--\cite{K2K}, which
implies neutrino mass and mixing.
In addition, the combined observations suggest that both the atmospheric
and solar mixing angles are nearly maximal, known as the large angle mixing
solution (LMA). Interestingly, the LMA solution
implies that the lepton mixing scenario is radically different
from the quark sector. From the low energy point of view,
we should expect that the neutrino mass inducing dimension 5
operator (HHLL) would be the first observable signal beyond the
renormalizable operators that compose the standard model. Furthermore,
the smallness of the inferred masses suggests
that the mechanism responsible for neutrino mass is distinct from
purely electroweak physics, and could naturally arise from
physics at a very high scale. The dimension 5 operator can be
induced by adding three heavy gauge singlet Majorana
fermions (one for each generation) to the
standard model. Upon integrating the heavy Majorana fermions out
at their associated scale, small neutrino masses are induced after electroweak symmetry breaking.
This neutrino mass generating technique is the see-saw mechanism \cite{see-saw}.

While the see-saw mechanism is an economical and natural way to
understand the smallness of the inferred neutrino masses, there are
many possible methods of implementing it, and therefore detailed neutrino
observations can be used to constrain GUT models. Perhaps the most
elegant GUT uses the grand unifying group SO(10) in four spacetime
dimensions. The spinor representation of SO(10) is 16 dimensional,
which accommodates all the helicity states of one fermion family
plus an extra singlet degree of freedom for a Majorana neutrino.
The generations are simply three copies of the spinor
representation. Since GUTs relate quark and lepton masses and
mixings, it is perplexing from a model building perspective as to
why lepton mixing is so different from that in the quark sector.
More specifically, it is of interest to understand why
$|U_{\mu3}|$ of the MNS matrix is so much larger than $|V_{cb}|$
of the CKM matrix. Over the last few years a number of models have
been developed to address this difference \cite{Albright1}--\cite{Ross}. Recently, a
particularly interesting and highly successful class of
supersymmetric SO(10) GUTs has emerged that makes use of
asymmetric mass matrices known as lopsided textures \cite{Albright1, Albright2, Albright3}. In these
models, the charged lepton sector is responsible for the large
atmospheric mixing angle while the Majorana singlet neutrino
matrix has a simple form that results in the large solar mixing
angle. Throughout this paper we will refer to these models as
the AB model class \cite{Albright1}.

After GUT breaking, these models reduce to the
R-parity conserving
minimal supersymmetric standard model (MSSM)
with specific model dependent relationships amongst the
Yukawa couplings. In addition to the constraints already provided
by the neutrino physics (and the demand that these models
reproduce all the low energy physics of the standard model), the
WMAP satellite observations \cite{WMAP} provide strong constraints on the
available supersymmetric parameter space if the lightest
supersymmetic particle (LSP) is assumed to compose the dark
matter \cite{Ellis:2003wm, Ellis:2003bm,susy-WMAP}.
Assuming the constraints on the CMSSM from the WMAP data,
the definite flavour structure of the AB models
will result in specific soft supersymmetry breaking parameters.
Therefore, the AB model class gives well defined predictions
for lepton flavour violation and in particular $\meg$.
It is of considerable interest to determine how the lepton flavour changing
neutral current bounds restrict the CMSSM parameters for the AB model class
in light of the WMAP data.

We organize this paper as follows. In section \ref{m_defn} we
outline the essential details of the AB models, the
supersymmetric parameter space, and the calculation for $\meg$.
We consider $\meg$ since at the present time, with the current bound \cite{muegam} of
$\mathrm{BR}(\meg) < 1.2 \times 10^{-11}$, this process gives
the strongest constraints on lepton flavour violation in the class
of models that we discuss. Furthermore, the MEG experiment at PSI \cite{meg}
expects to improve on this bound with the expected sensitivity of
$\mathrm{BR}(\meg) \lesssim 5 \times 10^{-14}$.
This experiment will provide stringent limits on models with
charged lepton flavour violation.
In section \ref{num_res} we display our
numerical results with the combined constraints from $\meg$ and
the WMAP satellite observations, and in section \ref{con} we
present our conclusions. The appendix provides further
calculational details.

\section{The AB Model Definition}
\label{m_defn}
The AB model class is based on an SO(10) GUT with a
U(1)$\times$Z$_2$$\times$Z$_2$
flavour symmetry and uses a minimum set of Higgs
fields to solve the doublet-triplet splitting problem \cite{Albright1, Albright2, Albright3}. The
interesting feature of these models is the use of a lopsided
texture. The approximate form of the charged lepton and the down
quark mass matrix in these models is given by
\be
\begin{array}{cc}
\Ye \sim \left( \begin{array}{ccc} 0 & 0& 0 \\
0& 0& \epsilon\\
0& \sigma&1
\end{array} \right),
&
\Yd \sim \left( \begin{array}{ccc} 0 & 0& 0 \\
0& 0& \sigma\\
0& \epsilon&1
\end{array} \right).
\end{array}
\ee
where $\sigma \sim 1$ and $\epsilon \ll 1$. As pointed out by the
authors of \cite{Albright1}, this asymmetric structure naturally occurs within a
minimal SU(5) GUT where the Yukawa interaction for the down quarks
and leptons is of the form $\lambda_{ij} {\bf \bar 5}_i {\bf 10}_j
{\bf 5_H}$ (${\bf 5_H}$ denotes the Higgs scalars). In an SU(5) GUT,
the left-handed leptons and the charge conjugate right-handed down
quarks belong to the ${\bf \bar 5}$ while the ${\bf 10}$ contains
the charge conjugate right-handed leptons and the left-handed down
quarks. Therefore the lepton and down quark mass matrices are
related to each other by a left-right transpose. Since SU(5) is a
subgroup of SO(10), this feature is retained in an SO(10) GUT.
This lopsided texture has the ability to explain why $|U_{\mu3}|
>> |V_{cb}|$. Making use of this observation, the AB models contain
the Dirac matrices ${\bf U},{\bf N},{\bf D}, {\bf L}$ for the
up-like quarks, Dirac neutrino interaction, down-like quarks, and
the leptons respectively \cite{Albright3},
\be
\begin{array} {cc}
{\bf U}= \left( \begin{array}{ccc} \eta & 0& 0 \\
0& 0& \epsilon/3\\
0& -\epsilon/3&1
\end{array} \right) M_U,
&
{\bf N}= \left( \begin{array}{ccc} \eta & 0& 0 \\
0& 0& -\epsilon\\
0& \epsilon&1
\end{array} \right)M_U,
\end{array}
\ee
\be
\begin{array}{cc}
{\bf D}= \left( \begin{array}{ccc} 0 & \delta& \delta^\prime e^{i\phi} \\
\delta & 0& \sigma + \epsilon/3\\
\delta^\prime e^{i\phi}& -\epsilon/3 &1
\end{array} \right)M_D,
&
{\bf L}= \left( \begin{array}{ccc} 0 & \delta& \delta^\prime e^{i\phi} \\
\delta & 0& -\epsilon\\
\delta^\prime e^{i\phi}& \sigma +\epsilon &1
\end{array} \right)M_D.
\end{array}
\ee
where
\be
\begin{array}{rclcrcl}
M_U &\approx& 113 \; \mathrm{GeV},
&\hspace{4mm}&
M_D &\approx& 1 \; \mathrm{GeV}, \\
\sigma &=& 1.78,
&\hspace{4mm}&
\epsilon &=& 0.145, \\
\delta &=& 8.6 \times 10^{-3},
&\hspace{4mm}&
\delta^\prime&=& 7.9 \times 10^{-3}, \\
\phi &=& 126^\circ,
&\hspace{4mm}&
\eta&=& 8 \times 10^{-6}.
\end{array}
\label{col-defs}
\ee
Dimensionless Yukawa couplings $\Yu, \Yn, \Yd$, and $\Ye$
can be extracted from the Dirac matrices.
The given values of $M_D$ and $M_U$
best fit the low energy data with $\tan \beta \approx 5$. It
should be noted that larger values of
$\tan \beta$ are easily accommodated by altering the values of $M_U$ and $M_D$ while
retaining accurate fits
to the low energy data after renormalization group running. The
lopsided texture of the AB model class nicely fits the large
atmospheric mixing angle; however, in order to obtain the large
solar mixing angle a specific hierarchical form of the heavy
Majorana singlet neutrino matrix needs to be chosen \cite{Albright2, Albright3}, namely,
\be
{\Mnu}= \left( \begin{array}{ccc} b^2\eta^2 & -b\epsilon\eta & a\eta \\
-b\epsilon\eta & \epsilon^2& -\epsilon\\
a\eta & -\epsilon &1
\end{array} \right)\Lambda_\mathrm{N}.
\label{right-handed}
\ee
where the parameters $\epsilon$ and $\eta$ are as defined in
equation (\ref{col-defs}). The parameters $a$ and $b$ are of order $1$ and
$\Lambda_\mathrm{N} \sim 2 \times 10^{14}$ GeV. Since the Majorana
singlet neutrino matrix is not related to the Dirac Yukawa
structure, it is not surprising that this matrix should take
on a form independent from the rest of the model. Once these
choices have been made, the AB model class is highly predictive and
accurately fits all the low energy standard model physics and the
neutrino mixing observations.

It should be emphasized that all these relations are defined at
the GUT scale and are therefore subject to renormalization group
running \cite{see-saw}. If we conservatively assume that
the GUT symmetry breaks directly to the standard model gauge symmetries,
SU(3)$\times$SU(2)$\times$U(1),
and that supersymmetry is broken super-gravitionally
through a hidden sector in a flavour independent manner, the AB
model class will give well defined predictions for charged lepton flavour violation.
There may also be significant contributions to the off-diagonal elements from
renormalization group running between the GUT and gravity scales \cite{Barb, Hisano0}.
Since the particulars of GUT and supersymmetry breaking -- as well as the
possibility of new physics above the GUT scale -- can have model dependent
effects on the branching ratio for $\meg$, we do not consider an
interval of running between the GUT and gravity scales.

The specific model predictions for the Dirac Yukawa couplings and
the form of the Majorana singlet neutrino matrix will feed into
the soft supersymmetry breaking slepton mass terms through
renormalization group running, generating off diagonal elements
that will contribute to flavour changing neutral currents
\cite{Borzumati:1986}.
The amount of flavour violation contained in the AB model class can be
examined through the branching ratio of the process $\meg$.

\section{Numerical Results for $\meg$}
\label{num_res}
After GUT and supersymmetry breaking, we have
the constrained minimal supersymmetric standard model (CMSSM)
with heavy gauge singlet neutrinos to make use of the see-saw mechanism.
The leptonic part of the superpotential is
\begin{equation}
\label{eq-W}
W = \epsilon_{\alpha\beta} H_\mathrm{d}^\alpha {\Eb}\Ye{\Lb}^\beta
  + \epsilon_{\alpha\beta} H_\mathrm{u}^\alpha {\Nb}\Yn{\Lb}^\beta
  + \frac{1}{2}{\Nb}\Mnu{\Nb}
\end{equation}
where $\Ye$, $\Yn$ are Yukawa matrices, and $\Mnu$ is the singlet Majorana
neutrino mass matrix.
The totally antisymmetric symbol is defined $\epsilon_{12}=+1$.
We explain our notation in detail in the appendix.
On integrating out the heavy singlet neutrinos,
equation (\ref{eq-W}) reduces to
\begin{equation}
\label{eq-Wss}
W =
\epsilon_{\alpha\beta} H_\mathrm{d}^\alpha \Eb\Ye\Lb^\beta
- \frac {1}{2} \nb^\mathrm{T} \mnu \nb
\end{equation}
where
\begin{equation}
\label{eq-mnulight}
\mnu = \frac{v^2}{2} \Yn^\mathrm{T} \Mnu^{-1} \Yn \sin^2\beta
\end{equation}
is the see-saw induced light neutrino mass matrix.
The coefficients $\beta$ and $v$ are defined in terms of Higgs fields expectation
values by
\begin{equation}
\frac{v^2}{2} =
\left< \Hdz \right>^2
+ \left< \Huz \right>^2
= \left (174\; \mathrm{GeV}\right)^2,
\quad \quad
\tan \beta = \frac{\left< \Huz \right>} {\left< \Hdz \right>}.
\end{equation}
The neutrino mass matrix, equation (\ref{eq-mnulight}),
is in general not diagonal and this
is the source of lepton flavour violating interactions.

We assume that supersymmetry is broken softly in that breaking occurs through  operators of mass dimension 2 and 3.
The soft supersymmetry breaking Lagrangian relevant to LFV studies is
\begin{eqnarray}
\la_\mathrm{breaking} &=&
-\delta_{\alpha\beta}
\tilde{\mathbf{L}}^{\alpha\dagger}
\mls \tilde{\mathbf{L}}^\beta
-\tilde{\Eb}\mes\tilde{\Eb}^\dagger
-\tilde{\Nb}\mns\tilde{\Nb}^\dagger \nonumber\\
& &
-\mhds \delta_{\alpha\beta} H_\mathrm{d}^{\alpha *} H_\mathrm{d}^\beta
-\mhus \delta_{\alpha\beta} H_\mathrm{u}^{\alpha *} H_\mathrm{u}^\beta
\nonumber\\
& &
\left(
-B\epsilon_{\alpha \beta} H_\mathrm{d}^\alpha H_\mathrm{u}^\beta
-\frac{1}{2}\tilde{\Nb}\mathbf{B}_{\tilde{\mathrm{N}}}\tilde{\Nb}
\cc\right)\nonumber\\
& &
\left(
-\epsilon_{\alpha\beta} H_\mathrm{d}^\alpha \tilde{\Eb}\Ae\tilde{\Lb}^\beta
-\epsilon_{\alpha\beta} H_\mathrm{u}^\alpha \tilde{\Nb}\An\tilde{\Lb}^\beta
\cc\right)\nonumber\\
\label{eq-SUSYbreak}
& &
\left(
- \frac{1}{2} M_1 \tilde{B} \tilde{B}
- \frac{1}{2} M_2 \tilde{W}^a \tilde{W}^a
\cc\right)
\end{eqnarray}
(see the appendix for the notational details).
The CMSSM assumes universal soft supersymmetry breaking parameters
at the supersymmetry breaking scale,
which we take to be of order the GUT scale, leading to the following GUT
relations:
\begin{eqnarray}
& &\mls = \mes = \mns = m_0^2 \cdot \unit, \\
& &\mhds = \mhus = m_0^2, \\
& &\Ae = \An = 0, \\
& &M_1 = M_2 = m_{1/2}
\end{eqnarray}
where $m_0$ and $m_{1/2}$ denote the
universal scalar mass and the universal gaugino mass respectively
($\unit$ is the 3$\times$3 unit matrix).
We conservatively assume that the trilinear terms $\Ae$ and $\An$
vanish at the supersymmetry breaking scale.

We run the parameters of the CMSSM using the renormalization group equations
(see appendix) working in a basis where the Majorana
neutrino singlet matrix is diagonal, integrating out each heavy neutrino
singlet at its associated scale.
After integrating down to the electroweak scale, we rotate the Yukawa couplings to the mass eigenbasis.
In order to understand the origin of flavour violation in this model class,
we first give a qualitative estimate.
The leading log approximation of the off-diagonal slepton mass term is
given by
\be\left(\Delta \mls\right)_{ij} \approx -
\frac{3}{8\pi^2}m_0^2({\bf Y_\nu}^\dagger {\bf Y_\nu})
\ln\left(\frac{M_\mathrm{GUT}}{{\Lambda_{\mathrm{N}}}}\right),
\ee
(assuming that the trilinears vanish at the GUT scale), and using this
approximation together with mass insertion techniques \cite{Hisano0, Hisano1}, the branching ratio for $\mu
\rightarrow e \gamma$ is
\bea
\mathrm{BR}(\mu \rightarrow e \gamma) &\sim&
\frac{\alpha^3}{G_\mathrm{F}^2}
\frac{\left(\left(\mls\right)_{12}\right)^2}{m_\mathrm{s}^8}
\tan^2 \beta \nonumber \\
&\approx& \frac{\alpha^3}{G_\mathrm{F}^2 m_\mathrm{s}^8} \left|
\frac{3}{8\pi^2}m_0^2\ln\frac{M_\mathrm{GUT}}{\Lambda_\mathrm{N}}\right|^2
\left|\left({\bf Y_\nu}^\dagger {\bf Y_\nu}\right)_{12}\right|^2 \tan^2 \beta
\label{tanbdepend}
\eea
where $m_s$ is a typical sparticle mass. We see that since the flavour structure of the AB
model class is specified so precisely, the branching ratio for $\meg$ is well determined.
In our calculation of the decay rate, we use the full one-loop expressions
derived from the diagrams in figure \ref{diags}
(see the appendix for more details).

The WMAP satellite observations \cite{WMAP} combined with constraints from $b \rightarrow s \gamma$ and
LEP direct searches \cite{PDG} strongly limit the available CMSSM parameter space if the LSP composes the dark matter
\cite{Ellis:2003wm, Ellis:2003bm,susy-WMAP}.
In addition to these constraints, realistic supersymmetric GUT models must also survive LFV bounds, such as the
limit on $\meg$. In particular, using all of the available bounds, both  cosmological and laboratory, we can further restrict the
AB model class.
\begin{figure}[ht!]
\begin{center}
\includegraphics[width=7cm]{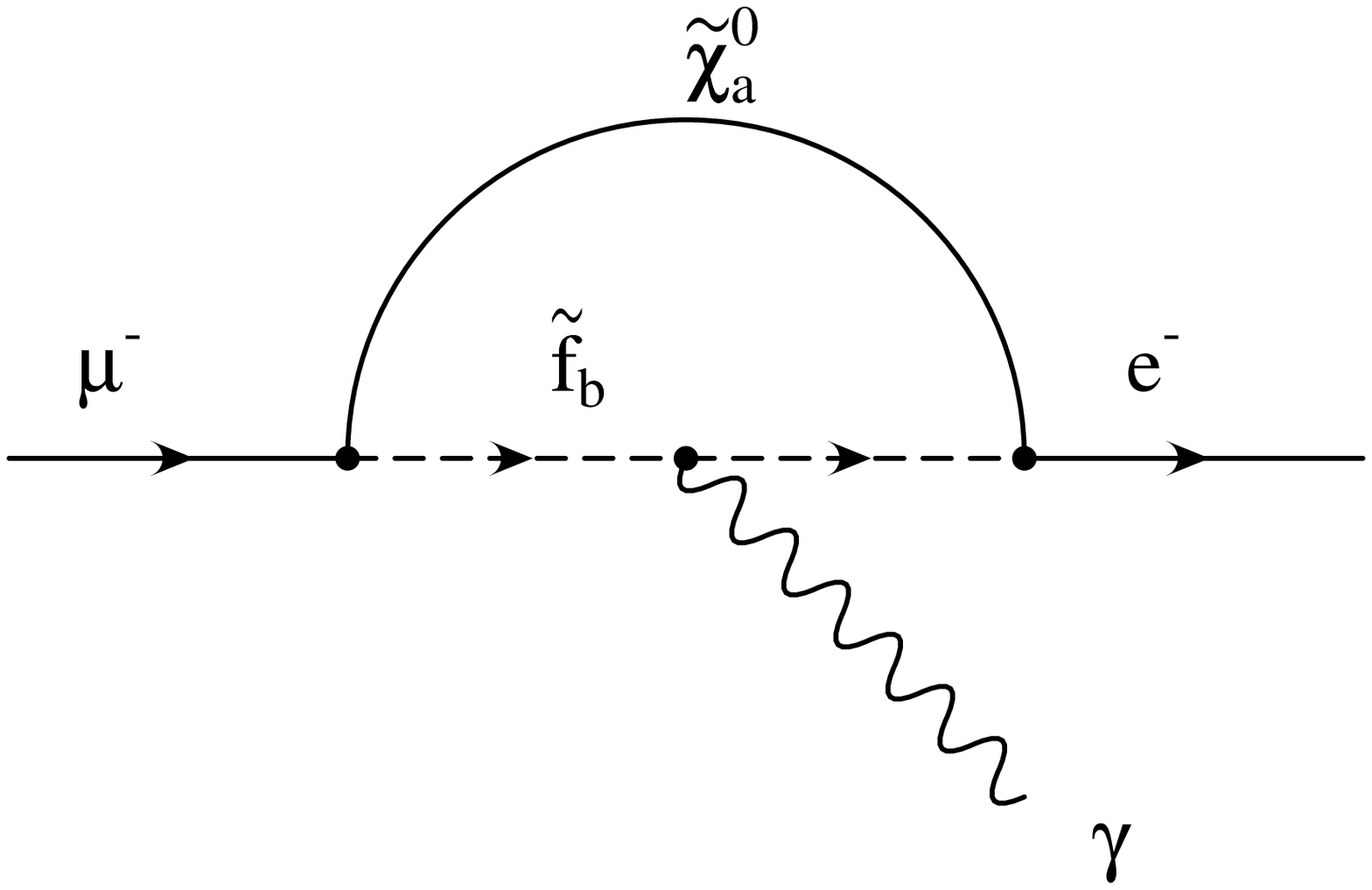}
\includegraphics[width=7cm]{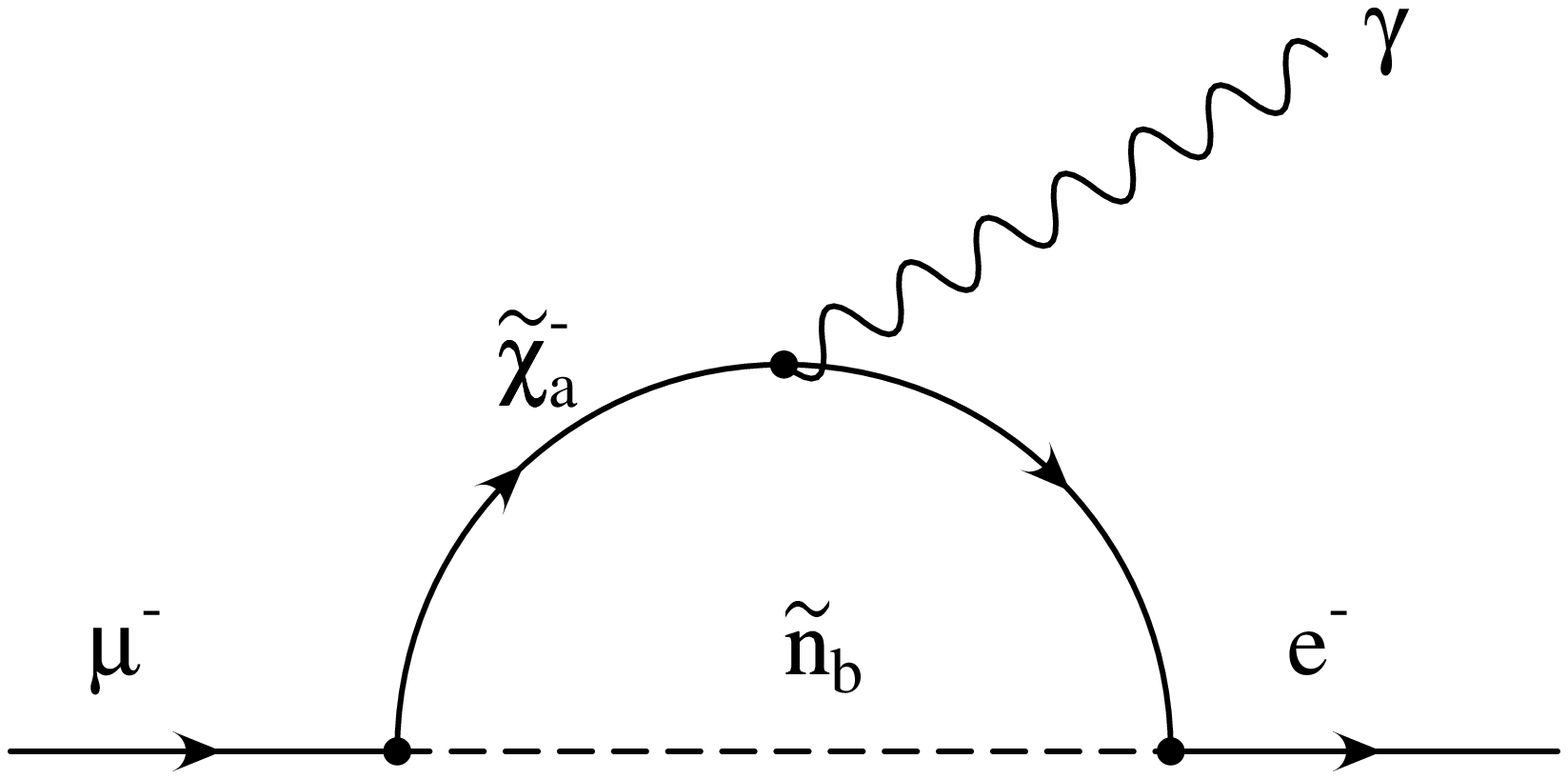}
\caption{Feynman diagrams contributing to $\mu\rightarrow e\gamma$.
\label{diags}}
\end{center}
\end{figure}

\begin{figure}[ht!]
\newlength{\picwidtha}
\setlength{\picwidtha}{2.1in}
 \begin{center}
\subfigure[][]{\resizebox{\picwidtha}{!}{\includegraphics{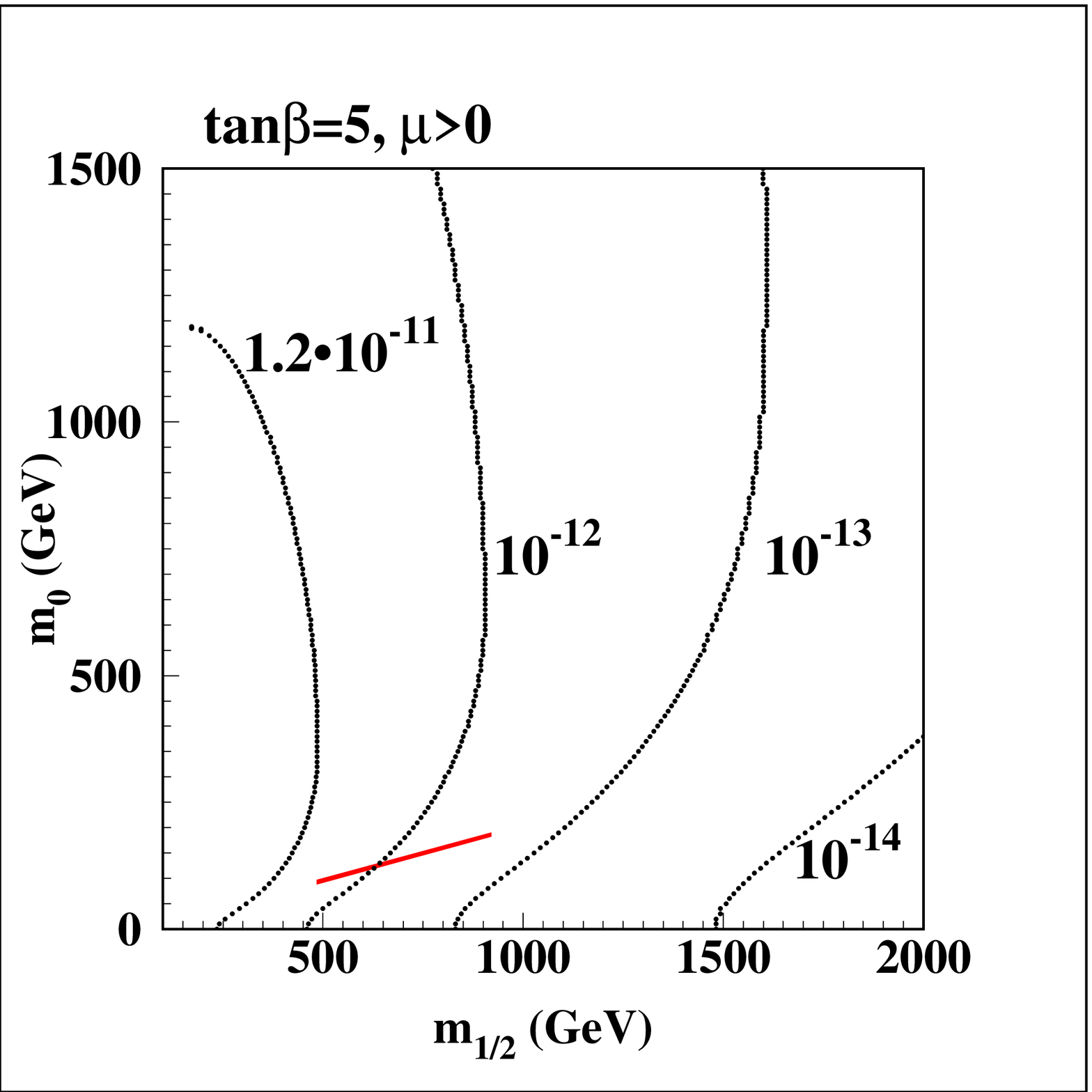}}}
 \subfigure[][]{\resizebox{\picwidtha}{!}{\includegraphics{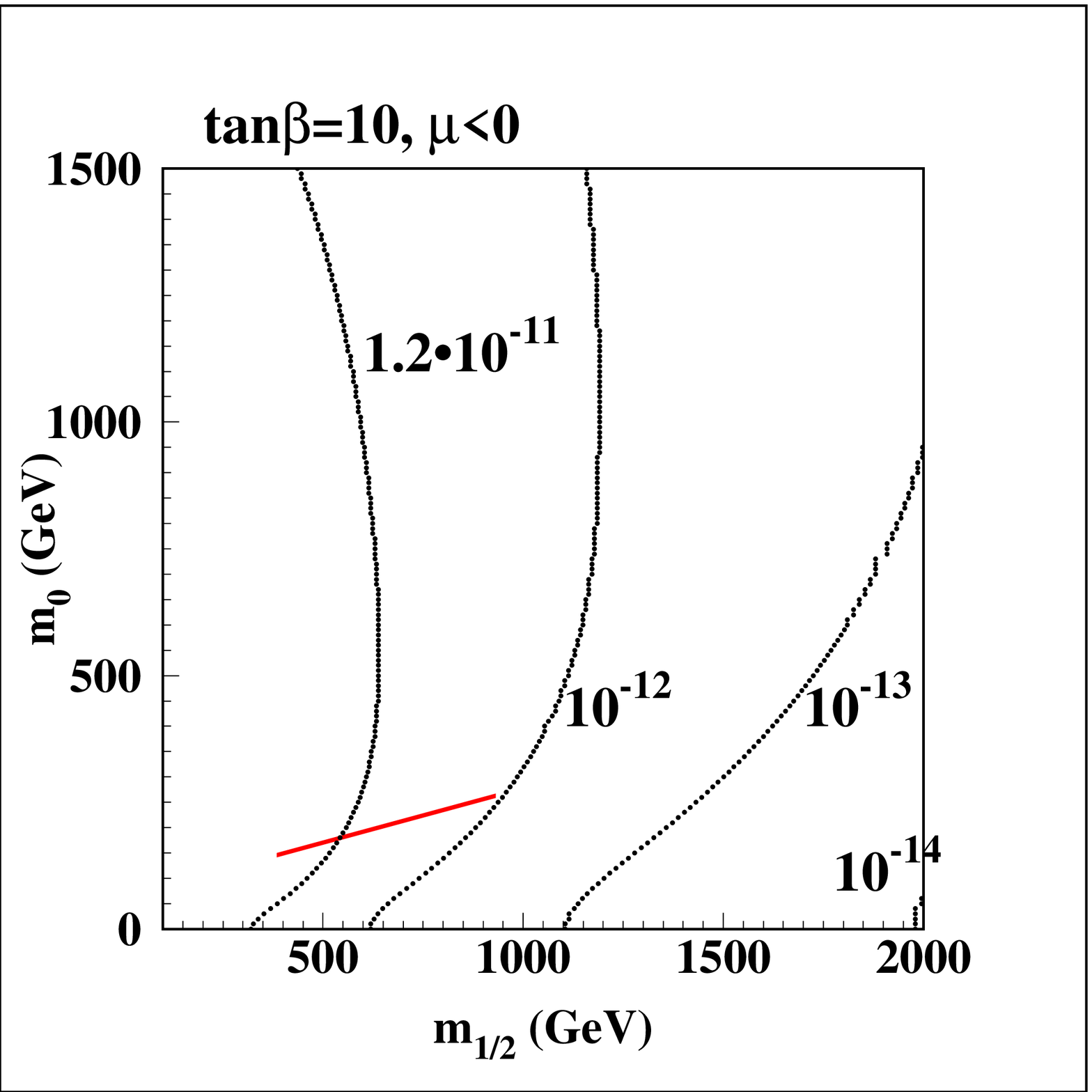}}}
 \subfigure[][]{\resizebox{\picwidtha}{!}{\includegraphics{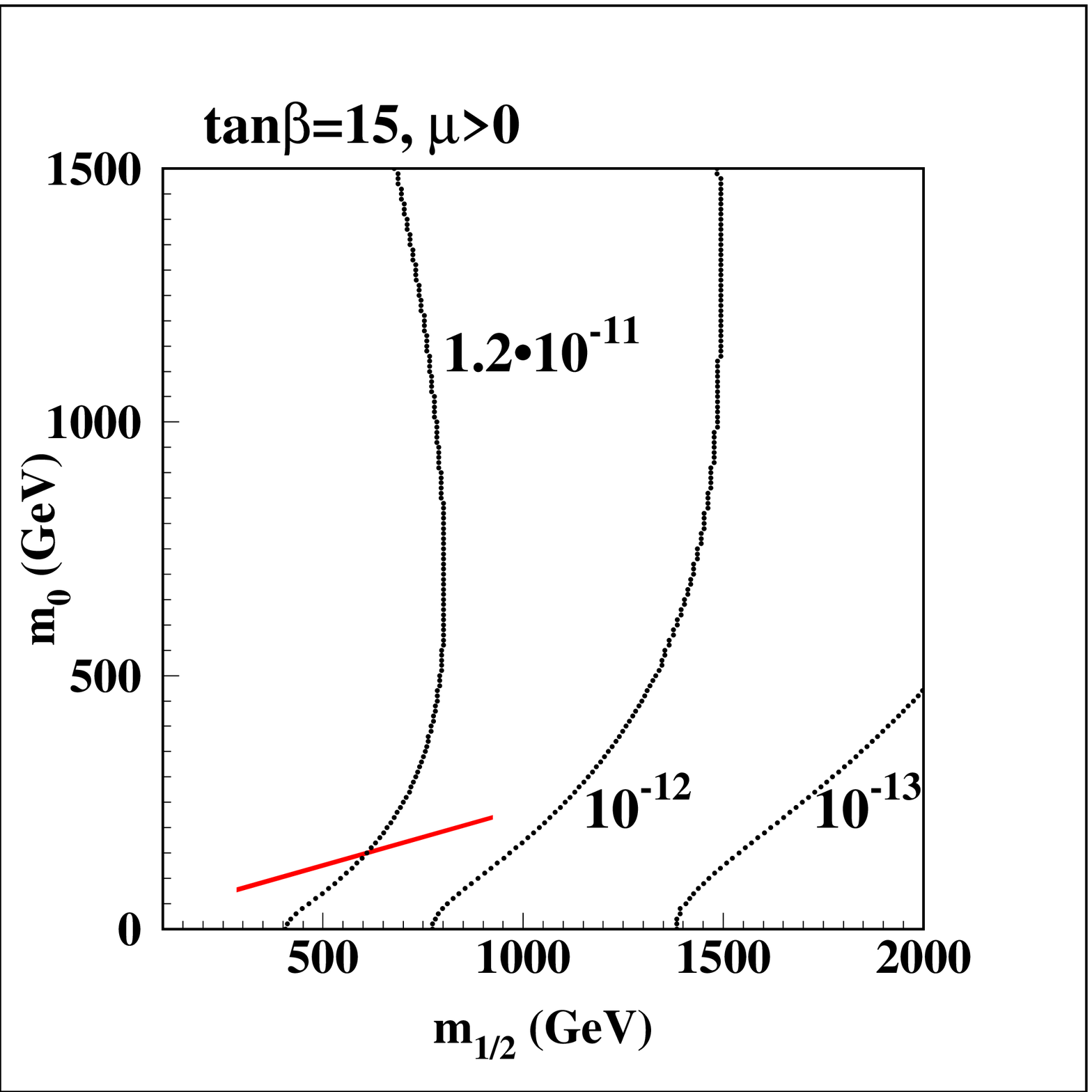}}}
 \subfigure[][]{\resizebox{\picwidtha}{!}{\includegraphics{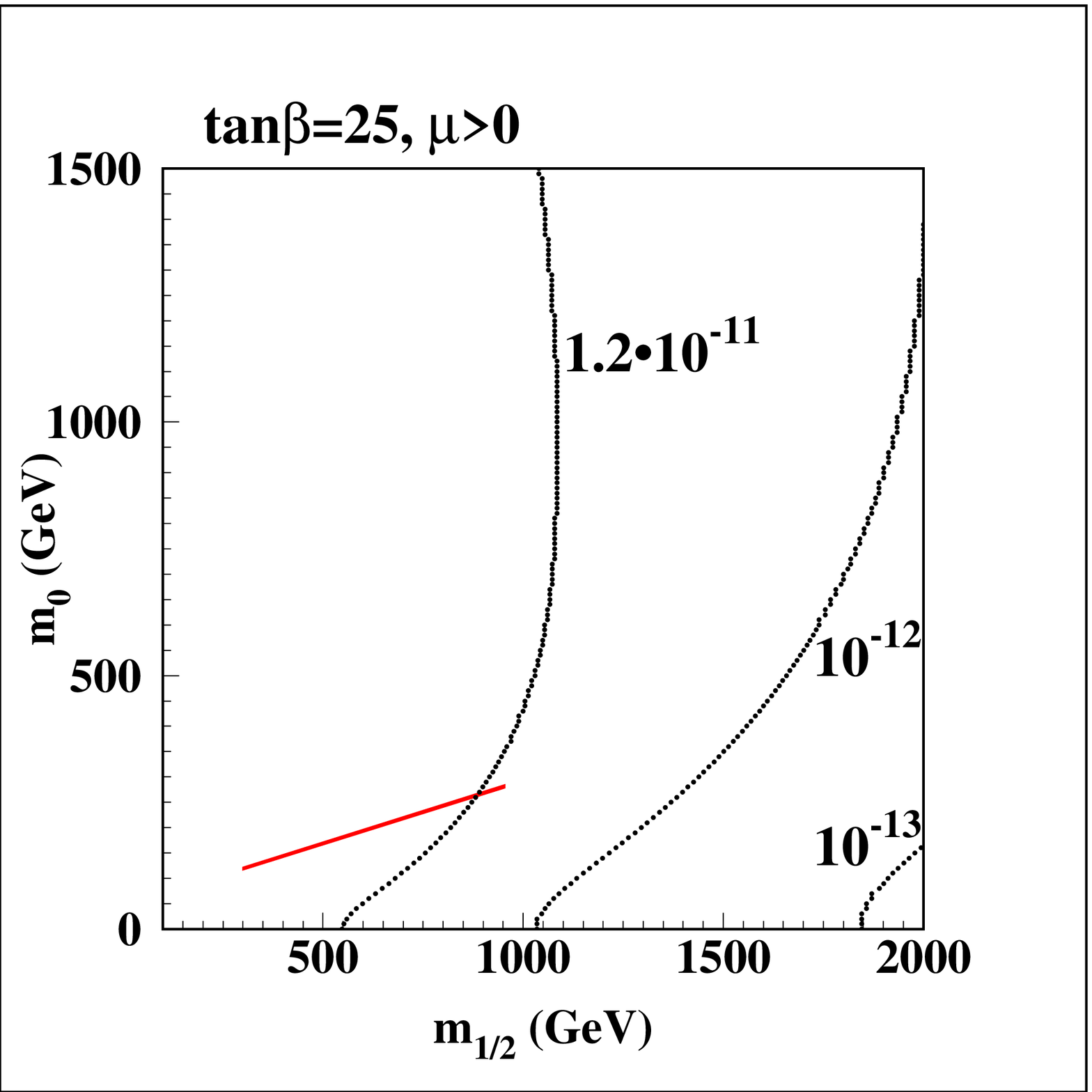}}}
  \subfigure[][]{\resizebox{\picwidtha}{!}{\includegraphics{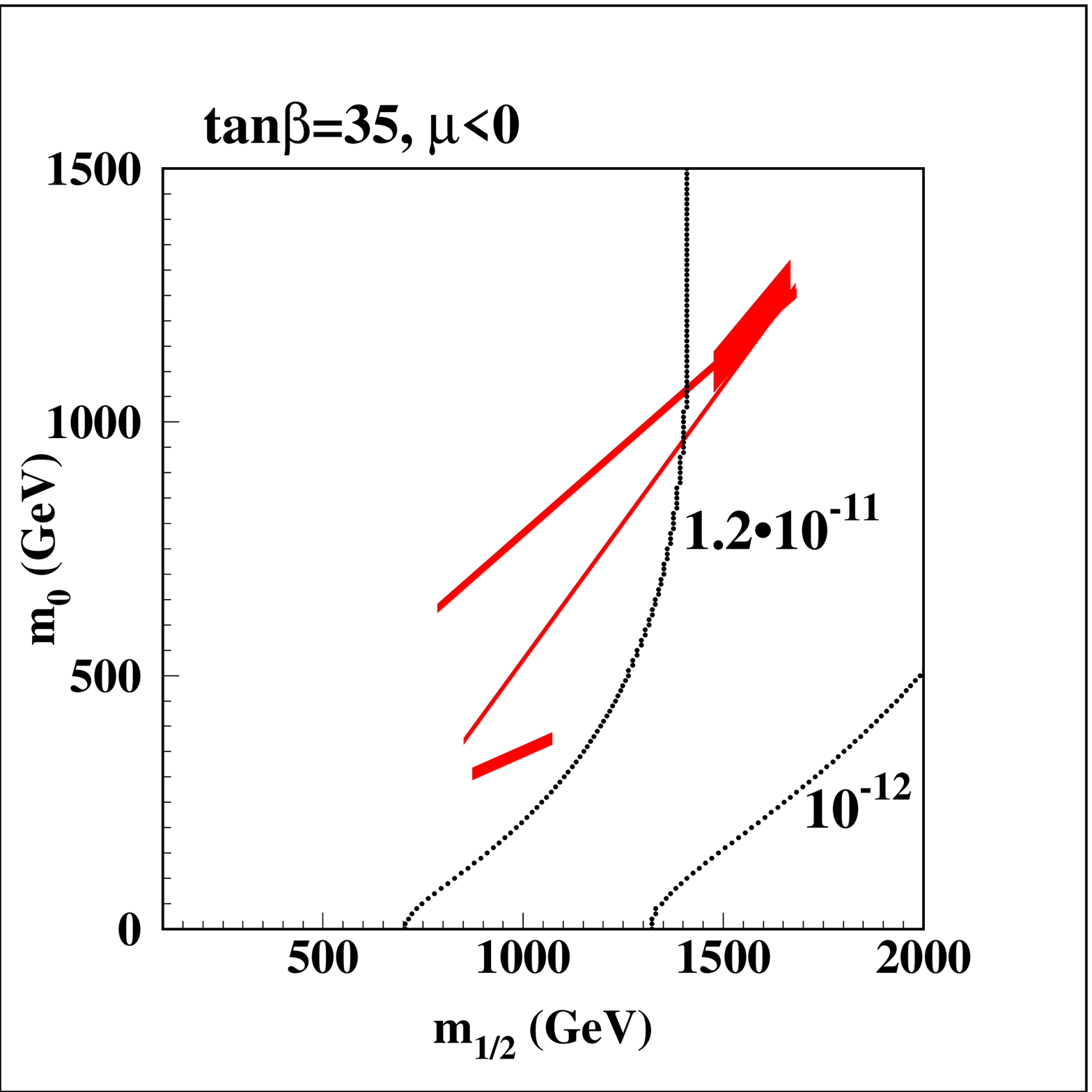}}}
   \subfigure[][]{\resizebox{\picwidtha}{!}{\includegraphics{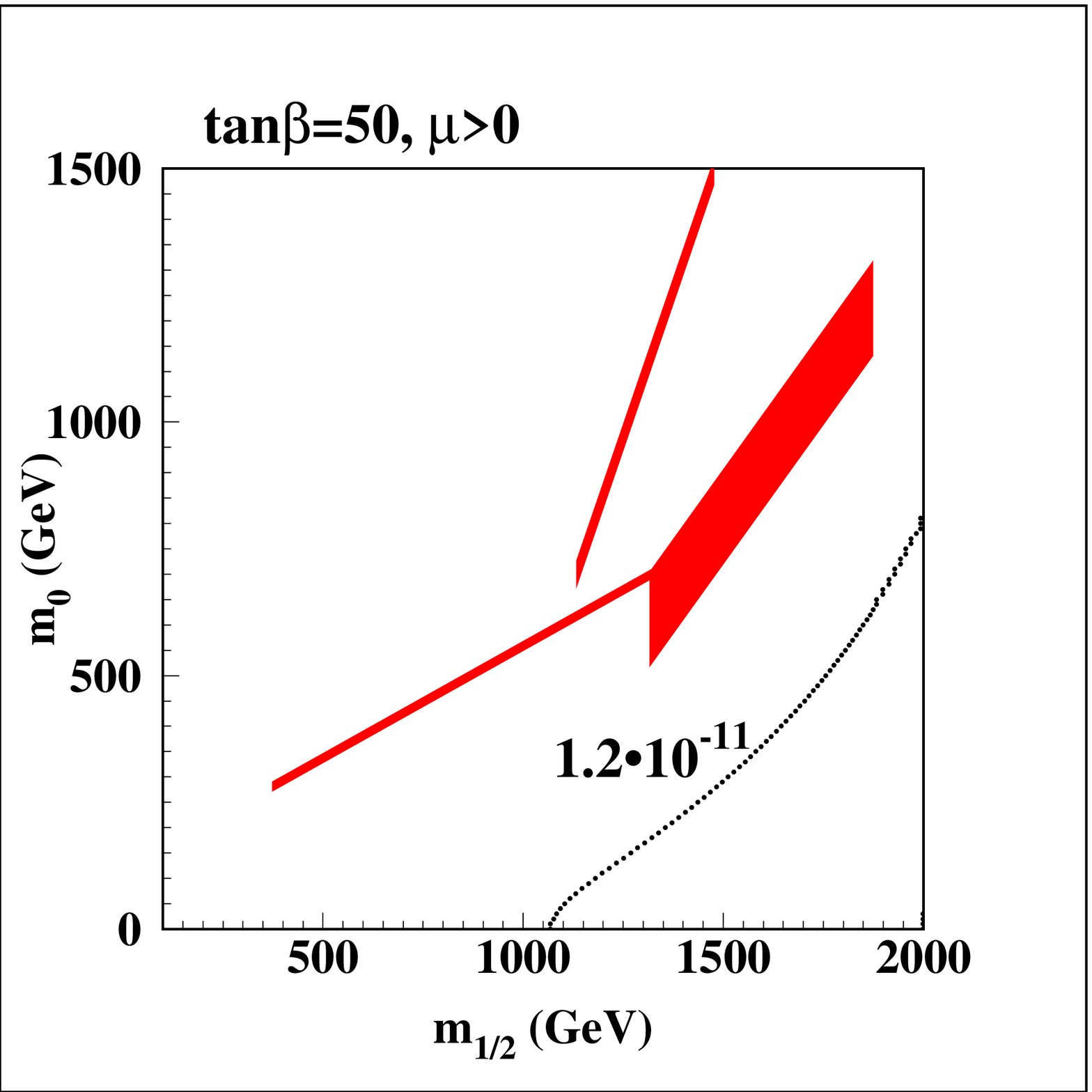}}}
 \end{center}
\caption{Contour Plots of $\mathrm{BR}(\meg)$ in the $m_0-m_{1/2}$ plane: Panels (a),(c),(d), and (f) show the contours of
the branching ratio for $\tan \beta= 5, 15, 25, 50$ respectively with $\mu>0$. Panels (b) and (e) show the contours with $\tan \beta=10,35$
respectively with $\mu <0$. In all cases the shaded region corresponds to the approximate combined WMAP and laboratory constraints. }
\label{emu}
\end{figure}

In figure \ref{emu}, we show contours of the branching ratio $\meg$ in the $m_{1/2}$-$m_{0}$ plane for a variety of $\tan \beta$ with the
$\mu$ parameter both positive and negative.
The parameters of the AB model class have been chosen such that all the low energy predictions fit the standard model data, and we have
chosen $a=1$ and $b=2$ for the Majorana singlet neutrino mass matrix
given in equation (\ref{right-handed}).
As indicated in \cite{Albright3}, there are a number
of possible model choices for the Majorana singlet parameters $a$ and $b$ that are consistent with the LMA solution. However, we find that the rate for
$\meg$ is largely unaffected by the allowed range \cite{Albright3}
for these parameters.
Panel (a) demonstrates the lepton flavour bounds for $\tan \beta =5$ with $\mu >0$ . The small line-like shaded area
in the lower part of the panel is the allowed region from the combined WMAP and laboratory limits.
The remaining panels show that the contours of constant branching ratio migrate to the right
of the plots (i.e. to high values of $m_{1/2}$ and $m_0$) as $\tan \beta$ is increased. In each case we overlay the
approximate WMAP and laboratory constraint bounds represented by a shaded region \cite{Ellis:2003wm}.
The choice for the sign of $\mu$ is indicated in each panel.
As $\tan \beta$ is pushed up, larger portions of the parameter space become
excluded.
This is an expected feature since the branching ratio is proportional to $\tan^2 \beta$. Notice that by $\tan \beta \sim 25$, $\mu>0$, the
branching ratio allowed contours no longer have a significant
overlap with the WMAP region.
As a result, we find that the  AB model class is consistent with the current
experimental bound on $\meg$ for low $\tan \beta$ (i.e. $\tan \beta \lesssim 20$) for $\mu >0$.
For completeness, in panels (b) and (e), we show two cases where $\mu<0$. The branching ratio of $\meg$ is largely insensitive to the sign of
$\mu$, however the WMAP region is moderately affected \cite{Ellis:2003bm}. A small part of the allowed WMAP region is currently permitted for larger $\tan \beta$
(i.e. $\sim 35$) as indicated in panel (e).  The upcoming limits \cite{meg} that MEG will establish,
$\mathrm{BR}(\meg) \lesssim 5 \times 10^{-14}$, will effectively rule out this
model class if LFV is not seen. Interestingly, if LFV is seen
at MEG, this model will suggest that $\tan \beta$ is low based on flavour bounds alone.
\section{Conclusions}
\label{con}
The AB model class \cite{Albright1, Albright2, Albright3}, based on a
U(1)$\times$Z$_2$$\times$Z$_2$
flavour symmetry, is a highly successful and predictive GUT scenario.
This model class has the ability to accommodate all the observed neutrino phenomena and reproduce the low energy physics of the standard model.
If it is assumed that supersymmetry is broken via mSUGRA and that the GUT breaks directly to the CMSSM,
the AB model class is highly restrictive and hence allows for a precise
determination for the rate of charged lepton flavour violation.
In particular, we examined the process $\meg$, since at the present time this flavour violating muon decay channel gives the
strongest constraints on flavour changing neutral currents in the lepton sector.

As the WMAP satellite data \cite{WMAP} and laboratory direct searches \cite{PDG} have already severely restricted the available CMSSM parameter space, the $\meg$
flavour bounds allow
a strong test of the AB model class. We find that given the current bounds \cite{muegam} on $\meg$, $\mathrm{BR}(\meg) < 1.2 \times 10^{-11}$, the AB model class favours
low $\tan \beta$ (i.e. $ \lesssim 20$) with $\mu >0$, however, there is a small region that is not excluded for $\tan \beta \lesssim 35$ with the sign of $\mu$ negative. If MEG at PSI \cite{meg} does not detect a positive LFV signal,
$\mathrm{BR}(\meg) \lesssim 5 \times 10^{-14}$, the AB model class will be effectively ruled out,
given our conservative assumptions concerning GUT and
supersymmetry breaking.  It remains an open question as to whether
or not other supersymmetry
and/or GUT breaking schemes within the AB model class will be able to avoid these flavour violating bounds.
\section{Acknowledgements}
We would like to thank Bruce A. Campbell for helpful discussions and comments.
DM acknowledges the support of the Natural
Sciences and Engineering Research Council of Canada
and EJ the support of the Alberta Ingenuity Fund.
\section{Appendix}
In this section we wish to clarify some of the calculational details. We carefully establish our notation and conventions.
Also, we include
the full one loop amplitude for the rate $\meg$ that we used
in our calculations.
Formulas similar to those given in subsections
\ref{neutralinos}-\ref{RGEs} can be found in \cite{Hisano1}.

We express the supersymmetric Lagrangian using the 2-component Weyl formalism.
$\Lb^\alpha = \left(L^\alpha_1,L^\alpha_2,L^\alpha_3\right)\T$
denotes a column vector in generation space containing the
SU(2) doublet lepton chiral superfields; 1,2,3 are generation labels, and
$\alpha=1,2$ are the SU(2) indices.
$\Eb= \left(E_1,E_2,E_3\right)$
denotes a row vector in generation space containing SU(2) singlet charged
lepton superfields.
The gauge singlet neutrino chiral superfields are denoted by
$\Nb= \left(N_1,N_2,N_3\right)$.
Similarly, for the quark superfields:
$\Qb^\alpha = \left(Q^\alpha_1,Q^\alpha_2,Q^\alpha_3\right)\T$
denotes the SU(2) doublet,
$\Qb^1=\ub=\left(u_1,u_2,u_3\right)\T$,
$\Qb^2=\db=\left(d_1,d_2,d_3\right)\T$;
and the SU(2) singlet quark superfields are
$\Ub= \left(U_1,U_2,U_3\right)$,
$\Db= \left(D_1,D_2,D_3\right)$.
$H_\mathrm{d}^\alpha$, $H_\mathrm{u}^\alpha$ are the SU(2) Higgs doublet
superfields of opposite hypercharge with the standard components:
$H_\mathrm{d}^{\alpha=1} = \Hdz$,
$H_\mathrm{d}^{\alpha=2} = \Hdm$,
$H_\mathrm{u}^{\alpha=1} = \Hup$,
$H_\mathrm{u}^{\alpha=2} = \Huz$.
The corresponding scalar components of the superfields are written
respectively as
$\tilde{\Lb}^\alpha$,
$\tilde{\Lb}^1 = \tilde{\nu}$,
$\tilde{\Lb}^2 = \tilde{\eb}$;
$\tilde{\Eb}$; $\tilde{\Nb}$;
$\tilde{\Qb}^\alpha$,
$\tilde{\Qb}^1 = \tilde{\ub}$,
$\tilde{\Qb}^2 = \tilde{\db}$;
$\tilde{\Db}$; $\tilde{\Ub}$;
(all are vectors in generation space).
The fermionic components of the Higgs superfield, the Higgsinos,
are denoted as
$\tilde{H}^\alpha_\mathrm{d}$,
$\tilde{H}^\alpha_\mathrm{u}$.
The superpotential $W$ is given by
\begin{eqnarray}
W = &\phantom{+}&
\epsilon_{\alpha\beta} H_\mathrm{d}^\alpha \Eb\Ye\Lb^\beta
+\epsilon_{\alpha\beta} H_\mathrm{u}^\alpha \Nb\Yn\Lb^\beta
+\frac{1}{2}\Nb\Mnu\Nb \nonumber\\
&+&\epsilon_{\alpha\beta} H_\mathrm{d}^\alpha \Db\Yd\Qb^\beta
+\epsilon_{\alpha\beta} H_\mathrm{u}^\alpha \Ub\Yu\Qb^\beta
\nonumber\\
\label{S_P}
&+&\mu \epsilon_{\alpha \beta} H_\mathrm{d}^\alpha H_\mathrm{u}^\beta
\end{eqnarray}
where $\Ye$, $\Yn$, $\Yd$, $\Yu$ are Yukawa matrices $\Mnu$ is
the singlet Majorana neutrino mass matrix,
$\mu$ is the Higgs parameter that
breaks the U(1) Pecci-Quinn symmetry,
and the totally antisymmetric symbol is defined $\epsilon_{12}=+1$.
The soft supersymmetry breaking Lagrangian is
\begin{eqnarray}
\la_\mathrm{breaking} &=&
-\delta_{\alpha\beta}
\tilde{\mathbf{L}}^{\alpha\dagger}
\mls \tilde{\mathbf{L}}^\beta
-\tilde{\Eb}\mes\tilde{\Eb}^\dagger
-\tilde{\Nb}\mns\tilde{\Nb}^\dagger \nonumber\\
& &
-\delta_{\alpha\beta}
\tilde{\mathbf{Q}}^{\alpha\dagger}
\mqs \tilde{\mathbf{Q}}^\beta
-\tilde{\Db}\mds\tilde{\Db}^\dagger
-\tilde{\Ub}\mus\tilde{\Ub}^\dagger \nonumber\\
& &
-\mhds \delta_{\alpha\beta} H_\mathrm{d}^{\alpha *} H_\mathrm{d}^\beta
-\mhus \delta_{\alpha\beta} H_\mathrm{u}^{\alpha *} H_\mathrm{u}^\beta
\nonumber\\
& &
\left(
-b\epsilon_{\alpha \beta} H_\mathrm{d}^\alpha H_\mathrm{u}^\beta
-\frac{1}{2}\tilde{\Nb}\mathbf{B}_{\mathrm{\tilde{N}}}\tilde{\Nb}
\cc\right)\nonumber\\
& &
\left(
-\epsilon_{\alpha\beta} H_\mathrm{d}^\alpha \tilde{\Eb}\Ae\tilde{\Lb}^\beta
-\epsilon_{\alpha\beta} H_\mathrm{u}^\alpha \tilde{\Nb}\An\tilde{\Lb}^\beta
\cc\right)\nonumber\\
& &
\left(
-\epsilon_{\alpha\beta} H_\mathrm{d}^\alpha \tilde{\Db}\Ad\tilde{\Qb}^\beta
-\epsilon_{\alpha\beta} H_\mathrm{u}^\alpha \tilde{\Ub}\Au\tilde{\Qb}^\beta
\cc\right)\nonumber\\
\label{eq-SUSYbreak}
& &
\left(
- \frac{1}{2} M_1 \tilde{B} \tilde{B}
- \frac{1}{2} M_2 \tilde{W}^a \tilde{W}^a
- \frac{1}{2} M_3 \tilde{G}^b \tilde{G}^b
\cc\right)
\end{eqnarray}
where
$\tilde{B}$ denotes electroweak U(1) gaugino field;
$\tilde{W}^a$, $a=1,2,3$, denote electroweak SU(2) gaugino fields;
$\tilde{G}^b$, $b=1,...,8$, denote strong interaction, SU(3),
gaugino fields;
$\mls$, $\mes$, $\mns$, $\mqs$, $\mds$, $\mus$,
$\mathbf{B}_\nu$, $\Ae$, $\An$, $\Ad$, $\Au$,
$\mhds$, $\mhus$, $b$, $M_1$, $M_2$, $M_3$
are the supersymmetry breaking parameters, and at the GUT scale:
\begin{eqnarray}
& &\mls = \mes = \mns = \mqs = \mds = \mus = m_0^2 \cdot \unit, \\
& &\mhds = \mhus = m_0^2, \\
& &\Ae = \An = \Ad = \Au = 0, \\
& &M_1 = M_2 = M_3 = m_{1/2}
\end{eqnarray}
where $m_0$ and $m_{1/2}$ denote the
universal scalar mass and the universal gaugino mass respectively ($\unit$ is the 3$\times$3 unit matrix). After running the CMSSM RGEs (see subsection \ref{RGEs}), we rotate all the Yukawa couplings to a diagonal basis, and in particular the lepton sector,
\begin{eqnarray}
\label{eq-r1}
\Ye \quad &\rightarrow& \quad \Ue^* \Ye \Ve^\dagger = \mathrm{diagonal},\\
\label{eq-r2}
\mls \quad &\rightarrow& \quad \Ve \mls \Ve^\dagger,\\
\label{eq-r3}
\mes \quad &\rightarrow& \quad \Ue^* \mes \Ue^T,\\
\label{eq-r4}
\Ae \quad &\rightarrow& \quad \Ue^* \Ae \Ve^\dagger.
\end{eqnarray}
Not all of the bi-unitary rotation matrices can be absorbed away
through the field
re-definitions as the left-handed neutrinos become massive below
the see-saw scale and after electroweak symmetry breaking.
\subsection{$\mu$ parameter}
The scalar potential of the Higgs fields is given at its minimum by
\begin{eqnarray}
V &=&
\left( \mu^2 + \mhds \right)
\left< \Hdz \right>^2
+\left( \mu^2 + \mhus \right)
\left< \Huz \right>^2 \nonumber\\
&+&\,b \left<\Hdz\right> \left<\Huz\right>
+ b^* \left<\Hdz\right> \left<\Huz\right> \nonumber\\
\label{eq-VHiggs}
&+&\,\frac{g_1^2 + g_2^2}{8}
\left(
\left< \Huz \right>^2 -  \left< \Hdz \right>^2
\right)^2
\end{eqnarray}
where $g_1$, $g_2$ are respectively U(1) and SU(2) gauge coupling constants.
We can use the SU(2) gauge transformation freedom to choose the vacuum
expectation value of the charged Higgs field $\left<\Hdm\right>=0$;
then it follows that also $\left<\Hup\right>=0$ at the minimum of the
Higgs potential. Therefore, we are left with only the neutral Higgs fields of
equation (\ref{eq-VHiggs}).
The conditions that the minimum of the potential $V$ breaks the electroweak
symmetry properly are
\begin{eqnarray}
\mu^2 + \mhds + b \tan \beta &=&
- \frac{1}{2} \mz^2 \cos 2\beta, \\
\mu^2 + \mhus + b \cot \beta &=&
\phantom{-} \frac{1}{2} \mz^2 \cos 2\beta
\end{eqnarray}
where $\mz$ is the mass of the $Z$-boson.
After eliminating the terms containing $b$ we obtain the tree level $\mu$ parameter relation,
\begin{equation}
\mu^2 =
-\half \mz^2
- \frac {\mhds - \mhus \tan^2 \beta}
{1-\tan^2 \beta}.
\end{equation}
\subsection{Neutralinos}
\label{neutralinos}
The neutralinos
$\tilde{\chi}^0_1$, $\tilde{\chi}^0_2$, $\tilde{\chi}^0_3$, $\tilde{\chi}^0_4$
are mass eigenstates of the neutral gauginos $\SB$, $\SWt$
and neutral Higgsinos $\SHdz$, $\SHuz$.
The neutralino mass Lagrangian is given by
\begin{equation}
\label{eq-laMneutralinos}
\la = -
\left(
\SB\; \SWt\; \SHdz\; \SHuz
\right)
\Mne
\left(
\begin{array}{c}
\SB \\
\SWt \\
\SHdz \\
\SHuz
\end{array}
\right)
\cc
\end{equation}
where
\begin{equation}
\label{eq-Mneutralinos}
\Mne =
\left(
\begin{array}{cccc}
M_1        & 0          & -\mz\cB\sw &  \phantom{-}\mz\sB\sw \\
0          & M_2        &  \phantom{-}\mz\cB\cw & -\mz\sB\cw \\
-\mz\cB\sw &  \phantom{-}\mz\cB\cw & \phantom{-}0 & -\mu \\
\phantom{-}\mz\sB\sw & -\mz\sB\cw & -\mu       & \phantom{-}0
\end{array}
\right).
\end{equation}
An orthonormal rotation leads to the mass eigenstates:
\begin{equation}
\left(
\begin{array}{c}
\tilde{\chi}^0_1 \\
\tilde{\chi}^0_2 \\
\tilde{\chi}^0_3 \\
\tilde{\chi}^0_4
\end{array}
\right)
= \ON
\left(
\begin{array}{c}
\SB \\
\tilde{W}^3 \\
\SHdz \\
\SHuz
\end{array}
\right)
\end{equation}
where $\ON$ is a real, orthogonal matrix.
The mass matrix (\ref{eq-Mneutralinos}) can therefore be decomposed in terms of real mass eigenvalues, $M_{\tilde{\chi}^0_a}$, $a=1,2,3,4$,
\begin{equation}
\mathbf{M}_\mathrm{ne} =
\ON^\mathrm{T}
\diag \left(
M_{\tilde{\chi}^0_1} \; M_{\tilde{\chi}^0_2} \;
M_{\tilde{\chi}^0_3} \; M_{\tilde{\chi}^0_4}
\right)
\ON,
\end{equation}
and (\ref{eq-laMneutralinos}) can be rewritten as
\begin{equation}
\la = -\frac{1}{2}
\sum_{a=1}^4
M_{\tilde{\chi}^0_a} \tilde{\chi}^0_a \tilde{\chi}^0_a.
\end{equation}
\subsection{Charginos}
The charginos are mass eigenstates of the charged SU(2) gauginos
and charged Higgsinos,
\begin{equation}
\label{eq-laMcharginos}
\la = - \left( \SWp \; \SHup \right)
\MC
\left(
\begin{array}{c}
\SWm \\
\SHdm
\end{array}
\right) \cc
\end{equation}
where
\begin{equation}
\tilde{W}^\pm = \frac {\tilde{W}^1 \mp i \tilde{W}^2} {\sqrt{2}},
\end{equation}
and the mass matrix is
\begin{equation}
\MC =
\left(
\begin{array}{cc}
M_2             & \sqrt{2} \mw \cB \\
\sqrt{2} \mw \sB & \mu
\end{array}
\right)
\end{equation}
($\mw$ is the $W$-boson mass). The mass eigenstates are given by
\begin{equation}
\left(
\begin{array}{c}
\tilde{\chi}^-_1 \\
\tilde{\chi}^-_2
\end{array}
\right)
= \OL
\left(
\begin{array}{c}
\SWm\\
\SHdm
\end{array}
\right),
\quad
\quad
%
%
\left(
\begin{array}{c}
\tilde{\chi}^+_1 \\
\tilde{\chi}^+_2
\end{array}
\right)
= \OR
\left(
\begin{array}{c}
\SWp\\
\SHup
\end{array}
\right)
\end{equation}
where $\OR$ and $\OL$ are real orthogonal matrices, and they can be chosen so
that the mass eigenvalues $M_{\tilde{\chi}^-_1}$, $M_{\tilde{\chi}^-_2}$
are positive, and
\begin{equation}
\MC = \OR^\mathrm{T}
\diag \left( M_{\tilde{\chi}^-_1} \; M_{\tilde{\chi}^-_2} \right)
\OL,
\end{equation}
Equation (\ref{eq-laMcharginos}) can be written as
\begin{equation}
\la =
-M_{\tilde{\chi}^-_1} \tilde{\chi}^+_1 \tilde{\chi}^-_1
-M_{\tilde{\chi}^-_2} \tilde{\chi}^+_2 \tilde{\chi}^-_2
\cc
\end{equation}
\subsection{Sleptons}
Masses of the charged sleptons are given by the Lagrangian
%
\begin{equation}
\la =
- \Se^\dagger \mLL \Se
- \Se^\dagger \mLR \SE^\dagger
- \SE \mRL \Se
- \SE \mRR \SE^\dagger
\end{equation}
with the mass matrices
\begin{eqnarray}
\mLL &=& \ml + \mls + \mz^2 \cos 2\beta
\left(\sin^2\theta_\mathrm{W} - \frac{1}{2}\right) \cdot\unit, \\
\mRR &=& \ml + \mes - \mz^2 \cos 2\beta \sin^2\theta_\mathrm{W} \cdot\unit,\\
\mRL &=& -\mu \mlfp \tan \beta + \frac{v \cB} {\sqrt{2}} \Ae
\end{eqnarray}
where
\begin{equation}
\mlfp = \diag \left(
m_{\mathrm{l}_1} \;
m_{\mathrm{l}_2} \;
m_{\mathrm{l}_3}
\right),
\end{equation}
and
$m_{\mathrm{l}_1}$,
$m_{\mathrm{l}_2}$,
$m_{\mathrm{l}_3}$
are electron, muon, and tau masses respectively.
The above Lagrangian written in terms of mass eigenstates
$\tilde{f}_1,...,$ $\tilde{f}_6$ (six complex scalar fields) is
\begin{equation}
\la =
- \sum_{b=1}^6
m^2_{\mathrm{\tilde{f}}_b} \tilde{f}_b^* \tilde{f}_b
\end{equation}
with
\begin{equation}
\left(
\begin{array}{c}
\tilde{f}_1 \\
\tilde{f}_2 \\
\tilde{f}_3 \\
\tilde{f}_4 \\
\tilde{f}_5 \\
\tilde{f}_6
\end{array}
\right)
= \Uf
\left(
\begin{array}{c}
\tilde{e}_1 \\
\tilde{e}_2 \\
\tilde{e}_3 \\
\tilde{E}_1^* \\
\tilde{E}_2^* \\
\tilde{E}_3^*
\end{array}
\right),
\end{equation}
and $\Uf$ is a complex unitary matrix defined by
\begin{equation}
\left(
\begin{array}{cc}
\mLL & \mLR \\
\mRL & \mRR
\end{array}
\right)
=
\Uf^\dagger
\diag
\left(
m^2_{\mathrm{\tilde{f}}_1}\;
m^2_{\mathrm{\tilde{f}}_2}\;
m^2_{\mathrm{\tilde{f}}_3}\;
m^2_{\mathrm{\tilde{f}}_4}\;
m^2_{\mathrm{\tilde{f}}_5}\;
m^2_{\mathrm{\tilde{f}}_6}
\right)
\Uf.
\end{equation}
Similarly, the light sneutrinos (the heavy singlet sneutrinos are ignored since they have decoupled well above the weak scale)
\begin{equation}
\la = - \tilde{\mathbf{\nu}}^\dagger \Msnu \tilde{\mathbf{\nu}}
\end{equation}
where
\begin{equation}
\Msnu =  \mls + \frac{1}{2} \mz^2 \cos 2\beta \cdot \unit.
\end{equation}
The sneutrino mass Lagrangian written in terms of mass eigenstates
$\tilde{n}_1$, $\tilde{n}_2$, $\tilde{n}_3$
(three complex scalar fields) reads
\begin{equation}
\la = -
\sum_{b=1}^3
m^2_{\mathrm{\tilde{n}}_b} \tilde{n}_b^* \tilde{n}_b
\end{equation}
with the mass eigenstates defined by
\begin{equation}
\left(
\begin{array}{c}
\tilde{n}_1 \\
\tilde{n}_2 \\
\tilde{n}_3
\end{array}
\right)
= \Un
\left(
\begin{array}{c}
\tilde{\nu}_1 \\
\tilde{\nu}_2 \\
\tilde{\nu}_3
\end{array}
\right),
\end{equation}
and $\Un$ is a complex unitary matrix satisfying
\begin{equation}
\Msnu =
\Un^\dagger
\diag
\left(
m^2_{\mathrm{\tilde{n}}_1}\;
m^2_{\mathrm{\tilde{n}}_2}\;
m^2_{\mathrm{\tilde{n}}_3}
\right)
\Un.
\end{equation}
\subsection{Lepton Flavour Violating Interactions}
The interactions leading to the lepton flavour violating process $l_j \rightarrow l_i + \gamma$
involve two effective Lagrangians: neutralino-lepton-slepton
and chargino-lepton-sneutrino.
Written in the mass eigenbasis they are
\begin{equation}
\label{eq-nls}
\la =
\sum_{i=1}^3 \sum_{a=1}^4 \sum_{b=1}^6
N^\mathrm{L}_{iab} \tilde{f}_b E_i \tilde{\chi}^0_a
+ N^{\mathrm{R}*}_{iab} \tilde{f}^*_b e_i \tilde{\chi}^0_a
\cc
\end{equation}
and
\begin{equation}
\label{eq-cls}
\la =
\sum_{i=1}^3 \sum_{a=1}^2 \sum_{b=1}^3
C^\mathrm{L}_{iab} \tilde{\nu}_b E_i \tilde{\chi}^-_a
+ C^{\mathrm{R}*}_{iab} \tilde{\nu}^*_b e_i \tilde{\chi}^+_a
\cc
\end{equation}
where
\begin{eqnarray}
\label{eq-NL}
N^\mathrm{L}_{iab} &=&
- \frac{g_2}{\sqrt{2}}
\left(
2\tan\theta_\mathrm{W}
\left(\Uf\right)^*_{b\left(i+3\right)}
\left(\ON\right)_{a1}
+ \frac{m_{\mathrm{l}_i}}{m_\mathrm{W}\cos\beta}
\left(\Uf\right)^*_{bi}
\left(\ON\right)_{a3}
\right),\\
N^\mathrm{R}_{iab} &=&
\frac{g_2}{\sqrt{2}}
\left(
\tan\theta_\mathrm{W}
\left(\Uf\right)^*_{bi}
\left(\ON\right)_{a1}
+ \left(\Uf\right)^*_{bi}
\left(\ON\right)_{a2}
- \frac{m_{\mathrm{l}_i}}{m_\mathrm{W}\cos\beta}
\left(\Uf\right)^*_{b\left(i+3\right)}
\left(\ON\right)_{a3}
\right),\quad \quad
\end{eqnarray}
and
\begin{eqnarray}
\label{eq-CL}
C^\mathrm{L}_{iab} &=&
\frac {g_2 m_{\mathrm{l}_i}} {\sqrt{2} m_\mathrm{W} \cos \beta}
\left(\OL\right)_{a2}
\left(\Un\right)^*_{bi},\\
\label{eq-CR}
C^\mathrm{R}_{iab} &=&
-g_2
\left(\OR\right)_{a1}
\left(\Un\right)^*_{bi}.
\end{eqnarray}
The on-shell amplitude for $l_j \rightarrow l_i + \gamma$ can be written in the general form
\begin{equation}
\label{eq-ampl}
\mathcal{M} =
e \epsilon^*_\mu \bar{l}_i\left(p-q\right)
\left(i m_{\mathrm{l}_j} \sigma^{\mu\nu} q_\nu
\left( A_\mathrm{L} \mathrm{L} + A_\mathrm{R} \mathrm{R} \right)
\right)
l_j \left(p \right);
\end{equation}
here we have used Dirac spinors
$l_i\left(p-q\right)$ and $l_j\left(p\right)$
for the charged leptons $i$ and $j$ with momenta $p-q$ and $p$,
respectively;
$\mathrm{L}=\left(1-\gamma^5\right)/2$ and
$\mathrm{R}=\left(1+\gamma^5\right)/2$.
Each of the dipole coefficients $A_\mathrm{L}$ and $A_\mathrm{R}$
have contributions from the neutralino-lepton-slepton
and the chargino-lepton-sneutrino interaction, namely,
\begin{equation}
\label{eq-AL}
A_\mathrm{L} = A_\mathrm{L}^\mathrm{(n)} + A_\mathrm{L}^\mathrm{(c)},
\end{equation}
\begin{equation}
\label{eq-AR}
A_\mathrm{R} = A_\mathrm{R}^\mathrm{(n)} + A_\mathrm{R}^\mathrm{(c)}
\end{equation}
where
$A_\mathrm{L}^\mathrm{(n)}$,
$A_\mathrm{R}^\mathrm{(n)}$,
$A_\mathrm{L}^\mathrm{(c)}$,
$A_\mathrm{R}^\mathrm{(c)}$
can be evaluated from the Feynman diagrams in figure \ref{diags};
\begin{eqnarray}
\label{eq-ALn}
A_\mathrm{L}^\mathrm{(n)} & = &
\phantom{-}\frac {1} {32 \pi^2}
\sum_{a=1}^{4} \sum_{b=1}^{6}
\frac {1} {m^2_{\mathrm{\tilde{f}}_b}}
\left(
N^\mathrm{L}_{iab} N^{\mathrm{L}*}_{jab}
J_1 \left( \frac {M^2_{\tilde{\chi}^0_a}} {m^2_{\tilde{l}_b}} \right)
+N^\mathrm{L}_{iab} N^{\mathrm{R}*}_{jab}
\frac{ \left|M_{\tilde{\chi}^0_a}\right| } {m_{l_j}}
J_2 \left( \frac {M^2_{\tilde{\chi}^0_a}} {m^2_{\tilde{l}_b}} \right)
\right), \\
%
%
\label{eq-ALc}
A_\mathrm{L}^\mathrm{(c)} & = &
-\frac {1} {32\pi^2}
\sum_{a=1}^{2} \sum_{b=1}^{3}
\frac {1} {m^2_{\tilde{\nu}_b}}
\left(
C^\mathrm{L}_{iab} C^{\mathrm{L}*}_{jab}
J_3 \left( \frac {M^2_{\tilde{\chi}^-_a}} {m^2_{\tilde{\nu}_b}} \right)
+C^\mathrm{L}_{iab} C^{\mathrm{R}*}_{jab}
\frac{ M_{\tilde{\chi}^-_a} } {m_{l_j}}
J_4 \left( \frac {M^2_{\tilde{\chi}^-_a}} {m^2_{\tilde{\nu}_b}} \right)
\right) \\
A_\mathrm{R}^\mathrm{(n)} &=&
\left.A_\mathrm{L}^\mathrm{(n)}\right|_{L \leftrightarrow R} \\
A_\mathrm{R}^\mathrm{(c)} &=&
\left.A_\mathrm{L}^\mathrm{(c)} \right|_{L \leftrightarrow R}.
\end{eqnarray}
The functions
$J_1\left(x\right)$,
$J_2\left(x\right)$,
$J_3\left(x\right)$,
$J_4\left(x\right)$
are defined as
\begin{eqnarray}
J_1\left(x\right) &=&
\frac
{1 - 6x + 3x^2 + 2x^3 - 6x^2 \ln x}
{6\left(1-x\right)^4},\\
J_2\left(x\right) &=&
\frac
{1 - x^2 + 2x\ln x}
{\left(1-x\right)^3},\\
J_3\left(x\right) &=&
\frac
{2 + 3x - 6x^2 + x^3 + 6x \ln x}
{6\left(1-x\right)^4},\\
J_4\left(x\right) &=&
\frac
{-3 + 4x - x^2 + 2\ln x}
{\left(1-x\right)^3}.
\end{eqnarray}
Finally, the decay rate for $l^-_j \rightarrow l^-_i + \gamma$ is given by
\begin{equation}
\label{eq-ratio}
\Gamma\left(l^-_j \rightarrow l^-_i + \gamma\right) =
\frac {e^2} {16\pi} m_{l_j}^5
\left(
\left|A_\mathrm{L}\right|^2
+\left|A_\mathrm{R}\right|^2
\right),
\end{equation}
and $i=1$, $j=2$ for $\mu \rightarrow e + \gamma$.
\subsection{Renormalization group equations (RGEs)}
\label{RGEs}
The general form of the supersymmetric renormalization group
equations \cite{Falck, Martin, Hisano1} are
\begin{equation}
\label{eq-X}
\frac{dX}{dt}=\frac{1}{16\pi^2}\dot{X}
\end{equation}
where $X$ is any of
$g_1$, $g_2$, $g_3$,
$\Yn$, $\Ye$, $\Yu$, $\Yd$,
$M_1$, $M_2$, $M_3$,
$\mhus$, $\mhds$,
$\mls$, $\mns$, $\mes$, $\mqs$, $\mus$, $\mds$,
$\An$, $\Ae$, $\Au$, $\Ad$, and the dotted quantities are listed below:
\begin{eqnarray}
\label{eq-g1}
\dot g_1 &=& 11 g_1^3, \\
\label{eq-g2}
\dot g_2 &=& g_2^3, \\
\label{eq-g3}
\dot g_3 &=& -3g_3^3, \\
\end{eqnarray}
%
%
\begin{equation}
\label{eq-Yn}
\dot\Yn=\Yn\left(
- g_1^2 \unit
- 3 g_2^2 \unit
+ 3\tr\left(\Yu^\dagger \Yu \right) \unit
+ \tr\left(\Yn^\dagger \Yn \right) \unit
+ 3 \Yn^\dagger \Yn
+ \Ye^\dagger \Ye
\right),
\end{equation}
\begin{equation}
\label{eq-Ye}
\dot\Ye=\Ye\left(
- 3 g_1^2 \unit
- 3 g_2^2 \unit
+ 3 \tr\left(\Yd^\dagger \Yd \right) \unit
+ \tr\left(\Ye^\dagger \Ye \right) \unit
+ 3 \Ye^\dagger \Ye
+ \Yn^\dagger \Yn
\right),
\end{equation}
\begin{eqnarray}
\label{eq-Yu}
\dot\Yu &=& \Yu\left(
- \frac{13}{9} g_1^2 \unit
- 3 g_2^2 \unit
- \frac{16}{3} g_3^2 \unit
+ 3 \tr\left(\Yu^\dagger \Yu \right) \unit
+ \tr\left(\Yn^\dagger \Yn \right) \unit \right. \nonumber\\
& & + \left. 3 \Yu^\dagger \Yu
+ \Yd^\dagger \Yd
\right),
\end{eqnarray}
\begin{eqnarray}
\label{eq-Yd}
\dot\Yd &=& \Yd\left(
- \frac{7}{9} g_1^2 \unit
- 3 g_2^2 \unit
- \frac{16}{3} g_3^2\unit
+ 3 \tr\left(\Yd^\dagger \Yd \right) \unit
+ \tr\left(\Ye^\dagger \Ye \right) \unit \right. \nonumber \\
& & + \left. 3 \Yd^\dagger \Yd
+ \Yu^\dagger \Yu
\right),
\end{eqnarray}
%
%
\begin{eqnarray}
\label{eq-M1}
\dot M_1 &=& 22 g_1^2 M_1, \\
\label{eq-M2}
\dot M_2 &=& 2 g_2^2 M_2, \\
\label{eq-M3}
\dot M_3 &=& -6 g_3^2 M_3,
\end{eqnarray}
\begin{equation}
\label{eq-S}
S=\mhus-\mhds+\tr\left(\mqs-2\mus+\mds-\mls+\mes\right),
\end{equation}
%
%
\begin{eqnarray}
\label{eq-mhu2}
\dot\mhus &=&
6 \tr\left(
  \mqs \Yu^\dagger \Yu
+ \Yu^\dagger \mus \Yu
+ \mhus \Yu^\dagger \Yu
+ \Au^\dagger \Au
\right) \nonumber \\
& & + 2 \tr\left(
  \mls \Yn^\dagger \Yn
+ \Yn^\dagger \mns \Yn
+ \mhus \Yn^\dagger \Yn
+ \An^\dagger \An
\right) \nonumber \\
& & -2g_1^2 M_1^2
- 6 g_2^2 M_2^2
+ g_1^2 S,
\end{eqnarray}
\begin{eqnarray}
\label{eq-mhd2}
\dot\mhds &=&
2 \tr\left(
  \mls \Ye^\dagger \Ye
+ \Ye^\dagger \mes \Ye
+ \mhds \Ye^\dagger \Ye
+ \Ae^\dagger \Ae
\right) \nonumber \\
& & + 6 \tr\left(
\mqs \Yd^\dagger \Yd
+ \Yd^\dagger \mds \Yd
+ \mhds \Yd^\dagger \Yd
+ \Ad^\dagger \Ad
\right) \nonumber \\
& & - 2 g_1^2 M_1^2
- 6 g_2^2 M_2^2 - g_1^2 S,
\end{eqnarray}
%
%
\begin{eqnarray}
\label{eq-ml2}
\dot\mls &=&
\mls \Ye^\dagger \Ye
+ \Ye^\dagger \Ye \mls
+ \mls \Yn^\dagger \Yn
+ \Yn^\dagger \Yn \mls \nonumber \\
& & + 2 \Ye^\dagger \mes \Ye
+ 2 \mhds \Ye^\dagger \Ye
+ 2 \Ae^\dagger \Ae \nonumber \\
& & + 2 \Yn^\dagger \mns \Yn
+ 2 \mhus \Yn^\dagger \Yn
+ 2 \An^\dagger \An \nonumber \\
& & - 2 g_1^2 M_1^2 \unit
- 6 g_2^2 M_2^2 \unit
- g_1^2 S \unit,
\end{eqnarray}
\begin{equation}
\label{eq-mn2}
\dot\mns =
2 \mns \Yn \Yn^\dagger
+ 2 \Yn \Yn^\dagger \mns
+ 4 \Yn \mls \Yn^\dagger
+ 4 \mhus \Yn \Yn^\dagger
+ 4 \An \An^\dagger,
\end{equation}
\begin{eqnarray}
\label{eq-me2}
\dot\mes &=&
2 \mes \Ye \Ye^\dagger
+ 2 \Ye \Ye^\dagger \mes
+ 4 \Ye \mls \Ye^\dagger
+ 4 \mhds \Ye \Ye^\dagger
+ 4 \Ae \Ae^\dagger \nonumber \\
& & - 8 g_1^2 M_1^2 \unit
+ 2 g_1^2 S \unit,
\end{eqnarray}
%
%
\begin{eqnarray}
\label{eq-mq2}
\dot\mqs &=&
\mqs \Yu^\dagger \Yu
+ \Yu^\dagger \Yu \mqs
+ 2 \Yu^\dagger \mus \Yu
+ 2 \mhus \Yu^\dagger \Yu
+ 2 \Au^\dagger \Au \nonumber \\
& & + \mqs \Yd^\dagger \Yd
+ \Yd^\dagger \Yd \mqs
+ 2 \Yd^\dagger \mds \Yd
+ 2 \mhds \Yd^\dagger \Yd
+ 2 \Ad^\dagger \Ad \nonumber \\
& & - \frac{2}{9} g_1^2 M_1^2 \unit
- 6 g_2^2 M_2^2 \unit
- \frac{32}{3} g_3^2 M_3^2 \unit
+ \frac{1}{3} g_1^2 S \unit,
\end{eqnarray}
\begin{eqnarray}
\label{eq-mu2}
\dot\mus &=&
2 \mus \Yu \Yu^\dagger
+ 2 \Yu \Yu^\dagger \mus
+ 4 \Yu \mqs \Yu^\dagger
+ 4 \mhus \Yu \Yu^\dagger
+ 4 \Au \Au^\dagger \nonumber \\
& & - \frac{32}{9} g_1^2 M_1^2 \unit
- \frac{32}{3} g_3^2 M_3^2 \unit
- \frac{4}{3} g_1^2 S \unit,
\end{eqnarray}
\begin{eqnarray}
\label{eq-md2}
\dot\mds &=&
2 \mds \Yd \Yd^\dagger
+ 2 \Yd \Yd^\dagger \mds
+ 4 \Yd \mqs \Yd^\dagger
+ 4 \mhds \Yd \Yd^\dagger
+ 4 \Ad \Ad^\dagger \nonumber \\
& & - \frac{8}{9} g_1^2 M_1^2 \unit
- \frac{32}{3} g_3^2 M_3^2 \unit
+ \frac{2}{3} g_1^2 S \unit,
\end{eqnarray}
%
%
\begin{eqnarray}
\label{eq-An}
\dot\An &=&
-g_1^2 \An
-3 g_2^2 \An
+3 \tr\left( \Yu^\dagger \Yu \right) \An
+ \tr\left( \Yn^\dagger \Yn \right) \An \nonumber \\
& & - 2 g_1^2 M_1 \Yn
- 6 g_2^2 M_2 \Yn
+ 6 \tr\left( \Yu^\dagger \Au \right) \Yn
+ 2 \tr\left( \Yn^\dagger \An \right) \Yn \nonumber \\
& & + 4 \Yn \Yn^\dagger \An
+ 5 \An \Yn^\dagger \Yn
+ 2 \Yn \Ye^\dagger \Ae
+ \An \Ye^\dagger \Ye,
\end{eqnarray}
\begin{eqnarray}
\label{eq-Ae}
\dot\Ae &=&
- 3 g_1^2 \Ae
- 3 g_2^2 \Ae
+ 3 \tr\left( \Yd^\dagger \Yd \right) \Ae
+ \tr\left( \Ye^\dagger \Ye \right) \Ae \nonumber\\
& & -6 g_1^2 M_1 \Ye
- 6 g_2^2 M_2 \Ye
+ 6 \tr\left( \Yd^\dagger \Ad \right) \Ye
+ 2 \tr\left( \Ye^\dagger \Ae \right) \Ye \nonumber\\
& & + 4 \Ye \Ye^\dagger \Ae
+ 5 \Ae \Ye^\dagger \Ye
+ 2 \Ye \Yn^\dagger \An
+ \Ae \Yn^\dagger \Yn,
\end{eqnarray}
\begin{eqnarray}
\label{eq-Au}
\dot\Au &=&
- \frac{13}{9} g_1^2 \Au
- 3 g_2^2 \Au
- \frac{16}{3} g_3^2 \Au
+ 3 \tr\left( \Yu^\dagger \Yu \right) \Au
+ \tr\left( \Yn^\dagger \Yn \right) \Au \nonumber\\
& & - \frac{26}{9} g_1^2 M_1 \Yu
- 6 g_2^2 M_2 \Yu
- \frac{32}{3} g_3^2 M_3 \Yu
+ 6 \tr\left( \Yu^\dagger \Au \right) \Yu
+ 2 \tr\left( \Yn^\dagger \An \right) \Yu \nonumber\\
& & + 4 \Yu \Yu^\dagger \Au
+ 5 \Au \Yu^\dagger \Yu
+ 2 \Yu \Yd^\dagger \Ad
+ \Au \Yd^\dagger \Yd,
\end{eqnarray}
\begin{eqnarray}
\label{eq-Ad}
\dot\Ad &=&
- \frac{7}{9} g_1^2 \Ad
- 3 g_2^2 \Ad
- \frac{16}{3} g_3^2 \Ad
+ 3 \tr\left( \Yd^\dagger \Yd \right) \Ad
+ \tr\left( \Ye^\dagger \Ye \right) \Ad \nonumber\\
& & - \frac{14}{9} g_1^2 M_1 \Yd
- 6 g_2^2 M_2 \Yd
- \frac{32}{3} g_3^2 M_3 \Yd
+ 6 \tr\left( \Yd^\dagger \Ad \right) \Yd
+ 2 \tr\left( \Ye^\dagger \Ae \right) \Yd \nonumber\\
& & + 4 \Yd \Yd^\dagger \Ad
+ 5 \Ad \Yd^\dagger \Yd
+ 2 \Yd \Yu^\dagger \Au
+ \Ad \Yu^\dagger \Yu.
\end{eqnarray}

\end{document}